\documentclass[aps,prl,superscriptaddress,twocolumn,floatfix,showpacs]{revtex4-1}

\usepackage{graphicx,graphics,epsfig}   
\usepackage{dcolumn}    
\usepackage{bm}         
\usepackage{amsmath}    
\usepackage{verbatim}   
\usepackage{color}      
\usepackage{subfigure}  
\usepackage{times,natbib}
\usepackage{amsmath,amsfonts,amssymb,graphics,graphicx,epsfig,color,times,natbib}

\newcommand{\be}{\begin{equation}}
\newcommand{\ee}{\end{equation}}
\newcommand{\ba}{\begin{eqnarray}}
\newcommand{\ea}{\end{eqnarray}}
\newcommand{\ban}{\begin{eqnarray*}}
\newcommand{\ean}{\end{eqnarray*}}

\newcommand{\ket}[1]{\mbox{$ | #1 \rangle $}}

\newcommand{\si}{\sigma}

\begin{document}

\title{Detecting Genuine Multipartite Quantum Nonlocality: \\A Simple Approach
and Generalization to Arbitrary Dimensions}

\author{Jean-Daniel Bancal}
\affiliation{Group of Applied Physics, University of Geneva, CH-1211 Geneva 4, Switzerland}
\author{Nicolas Brunner}
\affiliation{H.H. Wills Physics Laboratory, University of Bristol, Tyndall Avenue, Bristol, BS8 1TL, United Kingdom}
\author{Nicolas Gisin}
\affiliation{Group of Applied Physics, University of Geneva, CH-1211 Geneva 4, Switzerland}
\author{Yeong-Cherng Liang}
\affiliation{Group of Applied Physics, University of Geneva, CH-1211 Geneva 4, Switzerland}

\date{\today}

\begin{abstract}
The structure of Bell-type inequalities detecting genuine multipartite nonlocality, and hence detecting genuine multipartite entanglement, is investigated. We first present a simple and intuitive approach to Svetlichny's original inequality, which provides a clear understanding of its structure and of its violation in quantum mechanics. Based on this approach, we then derive a family of Bell-type inequalities for detecting genuine multipartite nonlocality in scenarios involving an arbitrary number of parties and systems of arbitrary dimension. Finally we discuss the tightness and quantum mechanical violations of these inequalities.
\end{abstract}


\maketitle

Non-locality is a fundamental feature of quantum mechanics. On top of being a fascinating phenomenon---defying intuition about space and time in a dramatic way---nonlocality is also a key resource for information processing \cite{resource}, and has thus been the subject of intense research in the last years.

It is fair to say that, while our comprehension of bipartite nonlocality has reached a reasonable level, multipartite nonlocality is still poorly understood. This is partly because the phenomenon becomes much more complex when moving from the bipartite case to the multipartite case. Indeed, this is somehow similar to the case of entanglement theory, where the structure of multipartite entanglement is much richer than that of bipartite entanglement \cite{genuine.multipartite}.

A natural issue to investigate is genuine multipartite nonlocality~\cite{svet87}, which represents the strongest form of multipartite nonlocality. More precisely, when considering a system composed of $m$  spatially  separated parts, it is natural to ask whether all $m$ parts of the system are nonlocally correlated, or whether it is only a subset of $k<m$ parts that display nonlocality while the remaining $m-k$ parts are simply classically correlated. Indeed such a question finds a natural context in quantum information theory and in the study of many-body systems \cite{manybody}. First, the presence of genuine multipartite nonlocality implies the presence of genuine multipartite entanglement. Also it is a fundamental issue to determine the role played by nonlocality in quantum information processing, for instance in measurement based quantum computation \cite{mbqc}.

In 1986, Svetlichny discovered the first method to detect genuine multipartite nonlocality \cite{svet87}. Focusing on the case of a system of three qubits, he derived a Bell-type inequality which holds even if (any) two out of the three parts would come together and act jointly---that is two parties can display arbitrary nonlocal correlations, while the third party is separated. A violation of such inequality implies that the systems features genuine tripartite nonlocality, implying the presence of  genuine  tripartite entanglement.
Svetlichny�s original inequality was later generalized to the case of an arbitrary number of parties~\cite{SvetGeneralized}, inspiring further studies on multipartite nonlocality in~\cite{mutinonlocality:furtherwork}. More refined concepts and measures of multipartite nonlocality have also been investigated \cite{multinonlocality:measure}.

In this Letter, we start by providing a simple and intuitive approach to Svetlichny's original inequality. Our approach, which naturally extends to the case of an arbitrary number of parties, makes it clear why these inequalities detects genuine multipartite nonlocality. It also provides an intuitive understanding of their violations in quantum mechanics, via the concept of steering~\cite{schroedinger35}. Based on this approach, we derive Bell inequalities detecting genuine multipartite nonlocality for an arbitrary number of systems of arbitrary dimension. Finally, we show that the simplest of our inequalities define facets of the relevant polytopes of correlations, and study their quantum mechanical violations. 

{\em Simple approach to Svetlichny's inequality} - To make the main idea of our approach clear, we first focus on the simplest scenario featuring three separated parties Alice, Bob, and Charlie. Each party  (labeled by $j$)  is asked to perform a measurement $X_j$ (chosen among a finite set) yielding a result $a_j$ with $j=1,2,3$. Thus the experiment is characterized by the joint probablity distribution $P(a_1a_2a_3|X_1X_2X_3)$. There exists different notions of nonlocality which the correlations $P$ can exhibit.

First, the experiment can display ``standard" nonlocal correlations, that is, the probability distribution $P$ cannot be written under the local form:
\ba\label{locality}  P_L(a_1a_2a_3) = \int d \lambda \rho(\lambda)P_1(a_1|\lambda)P_2(a_2|\lambda)P_3(a_3|\lambda) \ea
where $\lambda$ is a shared local variable and where we have omitted the measurement inputs $X_j$ for simplicity. To test for such type of nonlocality, one uses standard Bell inequalities.

However, this notion of nonlocality does not capture the idea of genuine multipartite nonlocality. For instance, in the case  where  Alice and Bob are nonlocally correlated, but uncorrelated from Charlie, it would still follow that $P$ cannot be written in the form \eqref{locality}, although the system features no genuine tripartite nonlocality.

To detect genuine multipartite nonlocality, one needs to ensure that the probability  distributions  cannot be reproduced by local means even if (any) two of the three parties would come together and act jointly---and consequently could reproduce any bipartite nonlocal probability distribution. Formally, this corresponds to  ensuring  that $P$ cannot be written  in  the form:
\ba\label{hybrid}  P_B(a_1a_2a_3) = \sum_{k=1}^{3} p_{k} \int d \lambda \rho_{ij}(\lambda)P_{ij}(a_ia_j|\lambda)P_k(a_k|\lambda) \ea
where  $\{i,j\}\bigcup\{k\}=\{1,2,3\}$,  and the sum takes care of different bipartitions of the parties. 
In the following we shall refer to such models as ``bipartition models". A probability distribution $P$ which cannot be expressed in the above form features genuine tripartite nonlocality; to be reproduced classically, all three parties must come together. Clearly, standard Bell inequalities can in general not be used to test for genuine multipartite nonlocality, and one needs better adapted tools.

From now on, we shall focus on the case where each party performs one out of two possible measurements. We denote the measurements of party $j$ by $X_j$ and $X'_j$, and their results by $a_j$ and $a'_j$. Considering the case of where $a_j,a'_j \in \{-1,1\}$, Svetlichny \cite{svet87} proved that the inequality
\ba\label{svet} S_3 &=& a_1a_2a'_3 + a_1a'_2a_3 + a'_1a_2a_3 - a'_1a'_2a'_3+ \\\nonumber  & & a'_1a'_2a_3 + a'_1a_2a'_3 + a_1a'_2a'_3 - a_1a_2a_3 \leq 4 \ea
holds for any probability distribution of the form \eqref{hybrid}. Thus a violation of inequality \eqref{svet} implies the presence of genuine tripartite nonlocality, and hence of genuine tripartite entanglement (regardless of the Hilbert space dimension \cite{bancal_prep}). The above polynomial should be understood as a sum of expectation values; for instance $a_1a_2a'_3$ means $E(a_1a_2a'_3)$, the expectation value of the product of the outcomes when the measurements are $X_1$, $X_2$, and $X'_3$.

We now start by rewriting inequality~\eqref{svet} as:
\ba\label{trick} S_3 &=& \text{CHSH}\, a'_3 + \text{CHSH}' \,a_3 \leq 4\ea
where $\text{CHSH}=a_1a_2+ a_1a'_2 + a'_1a_2 - a'_1a'_2$ is the usual Clauser-Horne-Shimony-Holt polynomial \cite{chsh}, and $\text{CHSH}'=a'_1a'_2+ a'_1a_2 + a_1a'_2 - a_1a_2$ is one of its  equivalent forms, obtained by inverting the primed and nonprimed measurements;
equivalently one could apply the mapping $a_1 \rightarrow a'_1$ and $a'_1 \rightarrow -a_1$. 

The main point of our observation is now the following: It is the input setting of Charlie that defines which version of the CHSH game Alice and Bob are playing. When C gets the input  $X'_3$,  then AB play the standard CHSH game; when C gets the input  $X_3$,  AB play CHSH'. From this observation, two simple arguments show that $S_3\leq4$ holds for any bipartition model of the form \eqref{hybrid}.
\begin{itemize}
  \item \emph{Argument 1.} Consider the bipartition AB/C. Although AB are together, and could thus produce any (bipartite) nonlocal probability distribution, they do not know which CHSH game they are supposed to play, as C is separated. Thus they are effectively playing the average game $\pm$CHSH$\pm$CHSH' (the signs specifying which game is played depend on the outputs of C). It can be immediately checked that the algebraic maximum of any of these average games is 4~\cite{fn:MaxAverage}. Hence, $S_3\leq4$ for the bipartition AB/C.

  \item \emph{Argument 2.} For the bipartition A/BC, B knows which version of the CHSH game he is supposed to play with A, since he is together with C. However, CHSH being a nonlocal game, AB cannot achieve better than the local bound (i.e. CHSH=2 or CHSH'=2), as they are separated~\cite{fn:OnC}. Thus it follows that $S_3\leq4$. Note that the same reasoning holds for the bipartition B/AC.
\end{itemize}

From  these arguments,  it follows that inequality \eqref{trick} holds for any correlation of the form \eqref{hybrid}. Note that since the polynomial $S_3$ is invariant under permutation of parties, the proof  already follows by applying either one of the two arguments given above. 
However, using both arguments above allows one in principle to deal with polynomials which are not  invariant   under permutation of parties.

Furthermore, expressing Svetlichny's inequality under the form \eqref{trick} allows one to understand its optimal quantum mechanical violation. Suppose $ABC$ share a three qubit Greenberger-Horne-
Zeilinger (GHZ) state $\ket{\psi}= (\ket{000}+\ket{111})/\sqrt{2}$. From \eqref{trick} it is clear that C should choose his measurement settings in order to prepare for AB the state that is optimal for the corresponding CHSH game, i.e. a maximally entangled state of two qubits. Let Alice and Bob choose measurements which are optimal for CHSH--- $X_1=\si_x$  and  $X'_1=\si_y$   for A;  $X_2=(\si_x -  \si_y)/\sqrt{2}$  and  $X'_2=(\si_x +  \si_y)\sqrt{2}$  for B. It is then straightforward to check that the measurements of C must be  $X_3=\si_x$  and  $X'_3=-\si_y$.   For instance, when C measures $\si_x$ and gets outcome $\pm1$, he prepares the state $\ket{\phi_{\pm}}=(\ket{00}\pm\ket{11})/\sqrt{2}$ for AB which is optimal for the CHSH game. Note that, given the measurement of A and B, the state $\ket{\phi_\pm}$ gives CHSH=$\pm2\sqrt{2}$; thus the output of C ensures that the overall sign is positive. Similarly, when C measures  $ - \si_y$  and gets outcome $\pm1$, he prepares for AB the state  $\ket{\tilde{\phi}_\pm}= (\ket{00}\pm i\ket{11})/\sqrt{2}$.  Given the measurements of A and B, the state $\ket{\tilde{\phi}_\pm}$ gives CHSH'=$\pm2\sqrt{2}$. Thus ABC achieve the score of $S_3=4\sqrt{2}$, which is the optimal quantum violation as can be checked using the techniques of Ref. \cite{npa}. Moreover, the idea of steering also allows one to understand the resistance to (white) noise of this quantum violation. Basically, Svetlichny's inequality should be violated if and only if the state of AB (prepared by a measurement of C) violates CHSH. Thus we expect the resistance to noise of the GHZ state for Svetlichny's inequality to coincide with the resistance to noise of a maximally entangled two qubit state for CHSH. Indeed, in both cases we get the critical visibility $w=1/\sqrt{2}$.

The form of inequality \eqref{trick} also suggests a straightforward generalization  to  an arbitrary number of parties $m$:
\ba\label{Sm} S_m = S_{m-1}\,a'_m + S'_{m-1}\,a_m \leq 2^{m-1}  \ea
where $S'_{m-1}$ is obtained from $S_{m-1}$ by applying the mapping $a_1 \rightarrow a'_1$ and $a'_1 \rightarrow -a_1$. From \emph{Argument 2} above it is clear that if inequality $S_{m-1}\leq 2^{m-2}$ holds for any bipartition of the $m-1$ parties, then inequality \eqref{Sm} holds for any bipartition where party $m$ is not alone. The fact that \eqref{Sm} holds for this partition as well follows from the fact that the polynomial $S_m$ is symmetric under permutation of the parties (see below).  Inequalities \eqref{Sm} are the generalizations of Svetlichny's inequality presented in Ref.~ \cite{SvetGeneralized}.

\emph{Detecting genuine multipartite nonlocality in systems of arbitrary dimension} --
The form \eqref{trick} suggests further generalizations. We now present a family of inequalities detecting genuine multipartite nonlocality for scenarios involving an arbitrary number of parties and systems of arbitrary dimension.  The main idea here consists of replacing  the CHSH expression in~\eqref{trick} with the Collins-Gisin-Linden-Massar-Popescu (CGLMP) expression \cite{CGLMP}, which  gives  bipartite Bell inequalities for systems of arbitrary dimension. Here we use the form of CGLMP introduced in Ref. \cite{Acin06}, that is
\ba\label{cglmp} S_{2,d} &=& [a_1+a_2]+[a_1+a'_2]^* \\\nonumber & & +[a'_1+a_2]^*+ [a'_1+a'_2-1] \geq d-1 \ea    
where $[X]=\sum_{j=0}^{d-1} j P(X=j \text{ mod }d)$ and $[X]^*=[-X]$. Note that for convenience the measurement outcomes are now denoted $a_j\in\{0,1,...,d-1\}$. Note also that for $d=2$, the CGLMP inequality reduces to CHSH.

To construct $S_{3,d}$ we use the idea of Eq. \eqref{trick}. First we define $S_{2,d}'$,  an equivalent form  of $S_{2,d}$ \cite{fn:rule} obtained using the rule:
\ba\label{rule} [...] \rightarrow [...+1]^* \text{    and    } [...]^*\rightarrow [...]. \ea
Next we construct  $S_{3,d}= S_{2,d}\circ a'_3 + S'_{2,d}\circ a_3 $  and obtain
\ba\label{S3d}\nonumber S_{3,d} &=& [a_1+a_2+a_3+1]^*+[a_1+a_2+a_3'] \\ & &+[a_1+a'_2+a_3] + [a'_1+a_2+a_3] \\\nonumber & &+  [a_1+a'_2+a'_3]^*+[a'_1+a_2+a'_3]^*
\\\nonumber && +[a'_1+a'_2+a_3]^*+ [a'_1+a'_2+a'_3-1]  \geq 2(d-1), \ea
 where the rule $\circ$ to include the  third party  works  by simply inserting its outcomes ($a_3$ or $a'_3$) into the brackets.  In the case $d=2$ this rule reduces to Eq.~\eqref{trick}. 

From the fact that $S_{2,d}$ is a Bell inequality and from \emph{Argument 2}, it follows that the inequality \eqref{S3d} holds for the bipartitions A/BC and B/AC. Moreover, since the polynomial $S_{3,d}$ is symmetric under permutation of the parties, the inequality \eqref{S3d} holds for any bipartition.

This construction can be generalized to an arbitrary number of parties $m$. Specifically, we take
\ba\label{Smd} S_{m,d}= S_{(m-1),d}\, a'_m + S'_{(m-1),d}\, a_m  \geq 2^{m-2}(d-1) \ea
where $S'_{(m-1),d}$ is obtained from $S_{(m-1),d}$ using the rule \eqref{rule}. For instance, for the case of $m=4$ parties we obtain
\ba\label{S4}\nonumber S_{4,d} &=& [a_1+a_2+a_3+a_4+1]+[a_1+a_2+a_3+a'_4+1]^* \\ && +[a_1+a_2+a'_3+a'_4] + [a_1+a'_2+a'_3+a'_4]^* \\\nonumber &&+ [a'_1+a'_2+a'_3+a'_4-1] +... \geq 4(d-1) \ea
where terms obtained by permuting the players are omitted.

\emph{Proof of inequality \eqref{Smd}.} -- The proof that \eqref{Smd} holds for any bipartition of the $m$ players is again based on \emph{Argument 2} and goes by induction. Let us suppose that (i) $S_{(m-1),d}\geq 2^{m-3}(d-1)$ holds for any bipartition of the $m-1$ parties and that (ii) $S_{(m-1),d}$ is invariant under any permutation of parties and contains all possible $2^{m-1}$ terms. Then, it follows from (i) that $S_{m,d}$ holds for all bipartitions, except for the one in which party $m$ is alone.

To deal with this last bipartition, we need to show that the polynomial $S_{m,d}$ is invariant under any permutation of parties. This is done in two steps. First note that by construction $S_{m,d}$ contains all $2^m$ possible terms. So it remains to be shown that all terms featuring a given number of unprimed inputs appear with the same type of brackets.
To see this, notice that the brackets associated with terms with an increasing number of unprimed measurements follow a regular pattern; terms featuring only primed measurements have $[...-1]$; terms with one unprimed measurement have $[...]^*$; terms with two unprimed measurements have $[...]$, etc. In order to determine the bracket of the following terms, one simply iterates the rule \eqref{rule}. So, the bracket of terms featuring $k$ unprimed measurements is obtained by starting from the bracket $[...-1]$ and iterating $k$ times the rule \eqref{rule}. Now, note that terms in $S_{m,d}$ featuring a fixed number of unprimed measurements $k$ can come from two possible terms: first, from terms in $S_{(m-1),d}$ featuring $k$ unprimed measurements; second from terms in $S'_{(m-1),d}$ featuring $k-1$ unprimed terms. From the pattern described above, it follows that both of these terms appear within exactly the same type of bracket. Thus we have that $S_{m,d}$ is symmetric under permutation of the parties, which completes the proof.  $\blacksquare$

 Note that the arguments presented above also allow us to construct $S_{m,d}$ directly using rule~\eqref{rule} starting from the bracket that contains only primed terms.  Moreover, it can be shown that $S_{m,2}$ is equivalent to the generalizations of Svetlichny's inequalities given in Ref.~\cite{SvetGeneralized}. 

\emph{Tightness.} -- Among Bell inequalities, those which define facets of the polytope of local correlations are of particular interest, since they form a minimal set of inequalities to characterize local correlations \cite{pitowsky}. These inequalities are referred to as ``tight" Bell inequalities. In this Letter, we  focus  on Bell-type inequalities detecting genuine multipartite nonlocality. These inequalities are thus satisfied by any bipartition model of the form \eqref{hybrid}. Indeed the set of bipartition correlations also forms a polytope---which is strictly larger than the local polytope~\cite{JD:Symmetric}. Here we  have checked  that inequalities \eqref{S3d} and \eqref{S4} are  facets  of the respective polytope for $d=2,3$.   We conjecture that all inequalities \eqref{Smd} correspond to facets.

\emph{Quantum violations.} -- Finally we discuss the quantum violation of our inequalities. In the case of Svetlichny's original inequality, it turned out that writing the inequality in the form \eqref{trick} naturally leads us to consider steering in order to find the optimal quantum violation. Indeed, since the structure of our inequalities \eqref{Smd} is based on \eqref{trick}, we follow a similar approach here, which will lead us to the optimal quantum violations as well.

First we recall that, in the bipartite case and for $d=3$, the maximal violation of the CGLMP inequality \eqref{cglmp} is obtained by performing measurements on a partially entangled state of two qutrits given by $\ket{\psi_2}=(\ket{00}+\gamma\ket{11}+\ket{22})/\sqrt{2+\gamma^2}$ where $\gamma=(\sqrt{11}-\sqrt{3})/2$ \cite{acin02}. The optimal measurements are so-called Fourier transform measurements \cite{CGLMP, Kaslikowski:PRL:00}; the basis is defined by the nondegenerate eigenvectors
$\ket{u} = \frac{1}{\sqrt{3}}\sum_{v=0}^2 \exp{\left[ \frac{2i\pi}{3}v(\alpha_m+u) \right]} \ket{v}$
for party $m$,  where $\alpha_1=0$, $\alpha'_1=-1/2$, and $\alpha_2=1/4$, $\alpha'_2=-1/4$ . This gives $S_{2,3}=1.0851$, corresponding to a resistance to (white) noise of $w=0.6861$.

Now moving to the case of three parties, it appears natural to choose the measurements of Alice and Bob to be the ones which are optimal for CGLMP  (i.e., as above). 
Next we choose the tripartite state and Charlie's measurements to be such that, by measuring his system, C prepares the desired state for A and B. For instance we can take simply $\ket{\psi_3}=\frac{1}{\sqrt{2+\gamma^2}}\left(\ket{000}+\gamma\ket{111}+\ket{222} \right)$
and fix Charlie's measurements to be Fourier transform as well---we take $\alpha_3=1/2$ and $\alpha'_3=0$. With these parameters we obtain the violation $S_{3,3}=2.1703$, which we  have checked   to be the optimal quantum violation using the techniques of Ref. \cite{npa}. Note also that the resistance to noise of $\ket{\psi_3}$ here is $w=0.6861$, which corresponds exactly to that obtained for CGLMP with $\ket{\psi_2}$.

From the structure of our inequalities \eqref{Smd}, we conjecture that this idea of steering always provides the optimal quantum violation, that is, that the optimal violation is always obtained from the state $\ket{\psi_m}=(\ket{0}^{\otimes m} +\gamma \ket{1}^{\otimes m} +\ket{2}^{\otimes m})/\sqrt{2+\gamma^2}$ and Fourier transform measurement. From this we expect the resistance to noise to be independent of the number of parties $m$ and given by $w=0.6861$. We could check numerically that this is indeed the case for $S_{4,3}$. Also, we expect a similar behavior for higher dimensions $d$.

\emph{Conclusion.} --
The main focus of this Letter is to provide an intuitive approach to Bell-type inequalities detecting genuine multipartite nonlocality. First, we provided a natural form for Svetlichny's inequality, which allows one to better understand its structure as well as its quantum violation. Based on this understanding, we then derived a family of Bell-type inequalities detecting genuine multipartite nonlocality for an arbitrary number of systems of arbitrary dimensionality. Finally, our approach suggests other possible generalizations. For instance it would be interesting to investigate the case where the parties can perform more than two measurements.

We thank D. Cavalcanti, S. Popescu, S. Pironio, O.
G\"uhne, and P. Skrzypczyk for insightful discussions. We also thank R. Chaves for pointing out an error in an earlier version of the manuscript. We acknowledge financial support from the UK EPSRC, the ERC-AG QORE, and the Swiss NCCR Quantum Photonics.

\emph{Note added.} -- Recently, we became aware of the work of Ref.~\cite{Chen2010} which presented an inequality sharing similar properties with our inequality Eq.~\eqref{Smd}.


\begin{thebibliography}{19}
\bibitem{fn:MaxAverage} This can be achieved by a local strategy between $AB$.
\bibitem{fn:OnC} Charlie�s output could in principle depend on Bob�s input. However, for inequalities involving only full correlations, such flip performed by $C$ can also be performed by $B$.
\bibitem{fn:rule} Instead of rule \eqref{rule}, $S'_{m,d}$ can also be obtained from $S_{m,d}$ via the relabelling: $a_1 \rightarrow a'_1$ and $a'_1 \rightarrow a_1+1$. Thus $S'_{m,d}$ is an equivalent form of $S_{m,d}$.
\end{thebibliography}
\end{document}